\documentclass[nofootinbib,showkeys,amssymb,amsmath,12pt,a4paper,titlepage]{revtex4}
\usepackage{amsfonts}
\usepackage{amsmath}
\usepackage{amssymb}
\usepackage{amsthm}
\usepackage{citesort}

\makeatletter
\def\@cite#1#2{\textsuperscript{[{#1\if@tempswa , #2\fi}]}}
\makeatother

\begin{document}

\title{\bf An analysis of a random algorithm for estimating all the matchings}

\author{Jinshan Zhang}
\email{zjs02@mails.tsinghua.edu.cn}
\author{Yan Huo}
\email{huoy03@mails.tsinghua.edu.cn}
\author{Fengshan Bai}
\email{fbai@math.tsinghua.edu.cn}

\affiliation{Department of Mathematical Sciences,\\ Tsinghua
  University, 100084,\\ Beijing, PRC.}

\begin{abstract}
  Counting the number of all the matchings on a bipartite graph has been transformed
  into calculating the permanent of a matrix obtained from the extended
  bipartite graph by Yan Huo, and Rasmussen presents a simple
  approach (RM) to approximate the permanent, which just yields a critical ratio
  O($n\omega(n)$) for almost all the 0-1 matrices, provided it's a simple promising
  practical way to compute this \#P-complete problem. In this paper,
  the performance of this method will be shown when it's applied to compute all the
  matchings based on that transformation. The critical ratio will
  be proved to be very large with a certain probability,
  owning an increasing factor larger than any polynomial of $n$ even in the sense for almost
  all the 0-1 matrices. Hence, RM fails to work well when counting all the matchings
  via computing the permanent of the matrix. In other words, we must
  carefully utilize the known methods of estimating the permanent
  to count all the matchings through that transformation.
\end{abstract}

\pacs{pacs numbers here}


\maketitle

\section[Introduction]{Introduction}

Let $G=(V,E)$ be a bipartite graph, where $V=V_1\cup V_2$ is the set
of vertices and $E\subset V_1\times V_2$ is the set of edges. In the
following sections we suppose $\#V_1=\#V_2=n$ if there's no special
illustration. A set of edges $S\subset E$ is called a matching if no
two distinct edges $e_1,e_2\in S$ contain a common vertex. $S$ is
called a k-matching if $\#S=k$. In special case, $S$ is called a
perfect matching if $k=n$. Let $S_k$ be the set of k-matching in $G$
and $A(G)$ be the set of all the k-matching, $k=0,1,\ldots,n$. For
the convenience of discussion, let $\#S_0=1$, then the number of all
the matchings in $G$ is $\#A(G)=\sum\limits_{i=0}^{n}\#S_k$.

The permanent of a 0-1 $A={a_{ij},1\leq i,j\leq n}$ is defined as
\begin{equation}
 Per(A)=\sum\limits_{\pi}\prod\limits_{i=1}^{n}a_{i,\pi(i)}
\end{equation}
where the sum is over all the permutations $\pi$ of
$[n]=\{1,\ldots,n\}$. It's well known that the permanent of an
adjacent matrix of bipartite graph equals the number of its perfect
matching. Let AM(G) denote the number of all the matchings in $G$,
and $A$ be adjacent matrix of $G$. \cite{Yhuo07} has proved that
\begin{equation}
AM(G)=\frac{1}{n!}per \left(\begin{array}{cc} A&I_{n \times n}\\1_{n
\times n}&1_{n \times n}\end{array}\right)
\end{equation}
where $I_{n \times n}$ is $n\times n$ unit matrix, $1_{n \times n}$
denotes  $n\times n$ matrix with all the elements 1. This means in
order to count the number of all the matchings of a bipartite graph
with $2n$ vertices we only need to compute the permanent of a
$2n\times 2n$ corresponding matrix transformed from adjacent matrix.
The computation of permanent has a long history and was shown to be
$\#$P-complete in \cite{Val79}. Thus, in the past $20$ years or so,
many random algorithms have been developed to approximate the
permanent, which can been divided at least four
categories\cite{Chien02}: elementary recursive algorithms(the
original one is Rasmussen method(RM)) \cite{Ras94}; reductions to
determinants \cite{Go81,Kar93,Bar99,Bar00}; iterative balancing
\cite{Lin00}; and Markov chain Monte Carlo \cite{Sin89,JSV01,KRS96}.
All these methods try to find a fully-polynomial randomized
approximation scheme $\emph{fpras}$ for computing the permanent.
$\emph{fpras}$ is such a scheme which, when given $\varepsilon$ and
inputs matrix $A$, outputs a estimator(usually a unbiased
estimator)$Y$ of the permanent such that
\begin{equation}
Pr((1-\varepsilon)per(A)\leq Y\leq(1+\varepsilon)per(A))\geq
\frac{3}{4}
\end{equation}
and runs in polynomial time in $n$ and $\varepsilon^{-1}$, here
$3/4$ may be boosted to $1-\delta$ for any desired $\delta>0$ by
running the algorithm $O(log(\delta^{-1}))$ and taking the median of
the trials \cite{Che52}. Then a straightforward application of
Chebychev's inequality shows that running the algorithm
$O(\frac{E(Y^2)}{E^2(Y)}\varepsilon^{-2})$ times and taking the mean
of the results can make the probability more than $3/4$(e.g. running
$4\frac{E(Y^2)}{E^2(Y)}\varepsilon^{-2}$ times). Hence, if the
critical ratio $\frac{E(Y^2)}{E^2(Y)}$ is bounded by a polynomial of
inputs $A$, we'll get an $\emph{fpras}$ for the permanent of $A$.
Another modified scheme called $\emph{fpras}$ for almost all inputs
means: choose a matrix from $\mathcal {A}(n,1/2)$($\mathcal
{A}(n,1/2)$ denotes a probability space of $n\times n$ 0-1 matrices
where each entry is chosen to be 1 or 0 with the same probability
1/2), or equivalently choose a matrix u.a.r. from $\mathcal {A}(n)$
($\mathcal {A}(n)$ represents the set of  $n\times n$ 0-1 matrices),
and the following

Pr(critical ratio of $A$ is bounded by a polynomial of the input $A$
)$=1-o(1)$ as $n\longrightarrow \infty$\\holds.(Note that this is a
much weaker requirement than that of an $\emph{fpras}$). If a
proposition $P$ relating to $n$ satisfies Pr(P is true)$=1-o(n)$, we
say P holds \textbf{whp}(\textbf{whp} is the abbreviation of "with
high probability"). Thus, that there is an $\emph{fpras}$ for almost
all the matrix means the critical ratio of $A$ is bounded by a
polynomial of the input $A$ \textbf{whp}. A exciting result, that
Markov Chain approach led to the first $\emph{fpras}$ for the
permanent of any 0-1 matrix(actually of any matrix with nonnegative
entry) was shown by\cite{JSV01}. However, its high exponent of
polynomial running time makes it difficult to be a practical method
to approximate the permanent. RM and reductions to determinants seem
to be two practical approaches estimating permanent due to their
simply feasibility, and both of them have been proved to be an
$\emph{fpras}$ for almost all the 0-1 matrices. besides,
\cite{Chien02} promises a good prospect on computing permanent via
clifford algebra if some difficulties can be conquered. RM also has
developed to be a kind of approaches called sequential importance
sampling way, which is widely used in statistical physics,
see\cite{BS99}.

In this paper, we'll, by RM, compute the number of all the matchings
based on the above transformation and give its performance
theoretically, say, an analysis of critical ratio in the sense "for
almost all the 0-1 matrix" of that matrix with a special structure.
In section II, A new alternative estimator operating directly on the
adjacent matrix without any transformation will be presented and
proved to be equivalent to approximation performing on the
transformed matrix by RM. In section III, a low bound of the
critical ratio for almost all the matrices will be presented, which
is larger than any polynomial of $n$ with a certain probability.
Hence, RM does not perform well in computing the number of all the
matchings as in computing the number of perfect matching. In section
IV we'll propose some analytic results w.r.t. the expectation and
variance of the number of all the matchings of a matrix selected
u.a.r from $\mathcal {G}(m,n)$($\mathcal {G}(m,n)$ denotes the set
of bipartite graph with $\#V_1=\#V_2=n$ as its vertices and exact
$m$ edges). These results seem likely to  contribute to the upper
bound of critical ratio for almost all matrices, but the
calculations are more arduous and will be left for latter paper.

\section {An equivalent estimator}

All the notations have the same meanings as those in the previous
section without special illustration. Let A an $n\times n$ 0-1
matrix be an adjacent matrix of a bipartite graph $G=(V,E)$,
$(V=V_1\bigcup V_2)$. Set $Y_A$ a random variable. Then RM
can be stated as follows:\\\\
\textbf{inputs:} A an $n\times n$ 0-1 matrix;\\
\textbf{outputs:} $Y_A$ the estimator of permanent A;\\
if n=0; then

$Y_A=1$ \\
else

   $W=\{j:a_{1j}=1\}$

   if $W=\emptyset$ then

      $Y_A=0$

   else

      Choose $J$ u.a.r. from W

      $Y_A=|W|Y_{1J}$\\\\
$Y_{1j}$ denotes the submatrix obtained from A by removing the 1st
row and the jth column. Note this heuristic idea comes from the
Laplace's expansion. Our following algorithm(for easy discussion,
call it AMM) is also inspired by another expansion. we first
presents our algorithm for the number of all the matchings, and then
give the explanation and proof of
equivalence between AMM and RM on the transformed matrix:\\\\
\textbf{inputs:} A an $n\times n$ 0-1 adjacent matrix of $G$;\\
\textbf{outputs:} $Y_A$ the estimator of the number of all the matchings of $G$;\\
if n=0; then

$Y_A=1$ \\
else

   $W=\{j:a_{1j}=1\}\bigcup\{0\}$

      Choose $J$ u.a.r. from W

      $Y_A=|W|Y_{1J}$\\\\
$Y_{10}$ denotes a submatrix of A by removing the 1st row(of course,
it's not necessarily a square matrix). Define a new terminology AM
on the matrix. let $B=\{b_{ij}, 1\leq i\leq m, 1\leq j\leq n\}$ an
$m\times n$ matrix, $m\leq n$. let $AM(\emptyset)=1$, by induction
on $m$.
\begin{equation}
AM(B):=AM(B_{10})+\sum\limits_{j=1}^{n}b_{1,j}B_{1j}
\end{equation}
Then we have the following theorem.\\\\
\textbf{Theorem 1.} Let A be an $n \times n$ adjacent matrix of a
bipartite graph G, Then AM(A)is the number of all the matchings of G.\\
\textbf{Proof:} It's easy to check, when $k\geq1$, the number of
k-matching of G equals
$\sum\limits_{i_1,\cdots,i_k}\sum\limits_{\pi}a_{i_1,\pi(i_1)}\cdots
a_{i_k,\pi(i_k)}$, where $i_1<i_2\cdots<i_k$ chosen from
$\{1,2,\cdots,n\}$, $\pi$ denotes the permutation
of$\{i_1,i_2,\cdots,i_k\}$. Thus, the number of all the matchings
is$\sum\limits_{k=1}^{n}\sum\limits_{i_1<\cdots<i_k\\\subseteq
\{1,\cdots,n\}}\sum\limits_{\pi}a_{i_1,\pi(i_1)}\cdots
a_{i_k,\pi(i_k)}+1$, where 1 denotes the number of 0-matching. Note
that if the AM(A) is written in terms of sum of elements of the
matrix A, then it's clearly to see
$AM(A)=\sum\limits_{k=1}^{n}\sum\limits_{i_1<\cdots<i_k\\\subseteq
\{1,\cdots,n\}}\sum\limits_{\pi}a_{i_1,\pi(i_1)}\cdots
a_{i_k,\pi(i_k)}+1$.$\Box$\\\\
 \textbf{Corollary1.} Let $A=\{a_{ij}1\leq
i,j\leq n\}$ be an $n\times n$ 0-1 matrix and $Y_A$ is obtained by
above AMM. Then
$Y_A$ is unbiased for AM(A), $E(Y_A)=AM(A)$\\
\textbf{Proof:} We prove for any $m\times n$ 0-1 matrix A, $1\leq
i\leq m, 1\leq j\leq n\}$, which will be widely used in the
following proves. AMM is unbiased for AM(A). For any fixed $n$, by
induction on m, k=0,$\forall 1\leq l\leq n$, $\forall$ a $k\times l$
0-1 matrix A, the equation $E(Y_A)=AM(A)$ is trivial. Now suppose
$\forall k\leq m, k\leq l\leq n$, a $k\times l$ 0-1 matrix A has
$E(Y_A)=AM(A)$. Then  when $k=m$, let $|W|=q$, we have

\begin{displaymath} \label{eq:1}
\begin{split}
E(Y_A)&= \sum\limits_{j\in W}E(Y_A|J=j)Pr(J=j)\\
      &= \sum\limits_{j\in W}E(qY_{A_{1j}}|J=j)q^{-1}\\
      &= \sum\limits_{j\in W}E(Y_{A_{1j}})\\
      &= \sum\limits_{j\in W}AM(A_{1j})\\
      &= AM(A).
 \end{split}
 \end{displaymath}$\Box$\\\\
Another simple corollary can also be obtained. To estimate the
number of all the matching in G, by RM operating on
$B=\left(\begin{array}{cc} A&I_{n \times n}\\1_{n \times n}&1_{n
\times n}\end{array}\right)$ divided by $n!$ is equivalent to
operating on A by AMM, in precise words, which can be stated as
follows.\\\\
\textbf{Corollary2.} Let $X_A$ be the output of RM operating on $A$
, $Y_B$ be the output of AMM operating on transformed matrix $B$
divided by $n!$. Then $X_A$ and $Y_B$ has the same
distribution.\\
\textbf{Proof:} Note that by RM after n-th step operating on $B$,
$Y_B=S_n*Y_{1_{n\times n}}/n!$, where $S_n$ is a number obtained
from the first n steps, and obviously $Y_{1_{n\times n}}\equiv n!$.
Hence, we have $Y_B=S_n$. The same distribution of $S_n$ and $X_A$
can be verified step by
step.$\Box$ \\\\
\textbf{Corollary3.} $AM(A)=\frac{1}{n!}per \left(\begin{array}{cc}
A&I_{n \times n}\\1_{n \times n}&1_{n \times
n}\end{array}\right)$.\\
\textbf{Proof:} This is a direct deduction of corollary2.  Let $X_A$
be the output of RM operating on $A$ , $Y_B$ be the output of AMM
operating on transformed matrix $B$ divided by $n!$.
\begin{displaymath}
AM(A)=E(X_A)=E(Y_B)=\frac{1}{n!}per \left(\begin{array}{cc} A&I_{n
\times n}\\1_{n \times n}&1_{n \times n}\end{array}\right)
\end{displaymath}$\Box$\\
So in the following section, we'll use AMM to compute all the
matchings instead of RM since some methodologies similar to
Rasmussen can be utilized. Another small advantage by AMM is that
the critical ratio is smaller than that directly obtained from RM.
The critical ratio by RM would be $(2n)!$, see Theorem
2.2\cite{Ras94}, while
the critical ratio by AMM would be $(n+1)^n$.\\\\
\textbf{Theorem2.} Let $A=\{a_{ij},1\leq i,j\leq n\}$ be an $n
\times n$ adjacent matrix of a bipartite graph G, and let $X_A$ be
the output of AMM. Then $\frac{E(X_A)^2}{E(X^{2}_{A})}\leq (n+1)^n$.
Generally, Let A be an $m\times n$ 0-1 matrix, $m\leq n$. $X_A$ be
the output of AMM. Then $\frac{E(X_A)^2}{E(X^{2}_{A})}\leq
(n+1)^m$\\
\textbf{Proof:} Induction on $m$, For any fixed $n$. $k=0$,$\forall
1\leq l\leq n$, $\forall$ a $k\times l$ 0-1 matrix $A$, the
inequation is trivial. In the case $k=m$, let $|W|=q$, we have
\begin{displaymath}
\begin{split}
E(X^2_A)&= \sum\limits_{j\in W}E(X^2_A|J=j)Pr(J=j)\\
        &= \sum\limits_{j\in W}E(q^2X^2_{A_{1j}}|J=j)q^{-1}\\
        &= \sum\limits_{j\in W}E(X^2_{A_{1j}})q\\
        &\leq \sum\limits_{j\in W}E(X_{A_{1j}})^2(n+1)^{m-1}q\\
        &\leq (\sum\limits_{j\in W}E(X_{A_{1j}}))^2(n+1)^{m-1}q\\
        &= E(X_A)^2(n+1)^{m}
\end{split}
\end{displaymath}$\Box$

\section {A lower bound of critical ratio for almost all the matrices}

Rasmussen shows that although the critical ratio of RM is factorial
in n, it does indeed provide an fpras for almost all the matrix.
However, the similar result can not be anticipated when computing
all the matchings by RM. In fact the critical ratio for almost all
the matrix would be more than $n^{\sqrt{n}/2-1}$ with a certain
probability. To prove this, we need to define some new denotations.
Since there're two probability spaces, we use the subscript $\sigma$
denote the calculus w.r.t. the probability space the algorithm lies
in, say, coin-tosses, and subscript $\mathcal {A}$ represent the
calculus w.r.t. the space probability the random matrices lie in.
$\mathcal {A}(m,n,p)$ denotes the probability space of all $m\times
n$ 0-1 random matrices where each entry is chosen to be 1 with
probability $p$, and$\mathcal {A}(m,n)$ denotes the set of all
$m\times n$ 0-1 matrices .

To obtain the mean and variance of the output of AMM on average
under probability measure $Pr_{\mathcal {A}}$, we need the following
lemma.\\\\
\textbf{Lemma1} Let $f(m,n)$ defined as
$f(m,n)=a_nf(m-1,n)+c_nf(m-1,n-1)$, where $m\leq n$ are two
nonnegative integers, $a_n$ and $c_n$ are two infinite positive
series w.r.t.
$n$. And $\forall$ $0\leq l\leq n$,  $f(0,l)=1$. Then\\
$f(m,n)=\sum\limits_{k=1}^{m}\sum\limits_{\substack{s_0+s_1+\cdots
+s_k=m-k \\s_0,\cdots s_k\geq 0}}c_n\cdots c_{n-k+1}a_{n}^{s_0}\cdots a_{n-k}^{s_k}+a_{n}^{m}$\\
\textbf{Proof:} By induction on m. Obviously, the case p=0 is
trivial. Suppose when $p\leq m-1$ $\forall$ $p\leq l\leq n$,
$f(p,l)=\sum\limits_{k=1}^{p}\sum\limits_{s_0+s_1+\cdots
+s_k=p-k}c_l\cdots c_{l-k+1}a_{l}^{s_0}\cdots
a_{l-k}^{s_k}+a_{l}^{p}$ holds, then when p=m, we have\\
\begin{displaymath}
\begin{split}
a_nf(m-1,n)&=\sum\limits_{k=1}^{m-1}\sum\limits_{s_0+s_1+\cdots
+s_k=m-1-k}c_n\cdots c_{n-k+1}a_{n}^{s_0+1}\cdots
a_{n-k}^{s_k}+a_{n}^{m}\\
           &=\sum\limits_{k=1}^{m-1}\sum\limits_{\substack{s_0+s_1+\cdots
+s_k=m-k\\ s_0\geq 1}}c_n\cdots c_{n-k+1}a_{n}^{s_0}\cdots
a_{n-k}^{s_k}+a_{n}^{m}
\end{split}
\end{displaymath}
and
\begin{displaymath}
\begin{split}
c_nf(m-1,n-1)&=\sum\limits_{k=1}^{m-1}\sum\limits_{s_0+s_1+\cdots
+s_k=m-1-k}c_n\cdots c_{n-k}a_{n-1}^{s_0}\cdots
a_{n-1-k}^{s_k}+c_na_{n-1}^{m-1}\\
             &=\sum\limits_{k=1}^{m-1}\sum\limits_{s_1+s_2+\cdots
+s_{k+1}=m-1-k}c_n\cdots c_{n-k}a_{n-1}^{s_1}\cdots
a_{n-1-k}^{s_{k+1}}+c_na_{n-1}^{m-1}\\
             &=\sum\limits_{k=2}^{m}\sum\limits_{s_1+s_2+\cdots
+s_{k}=m-k}c_n\cdots c_{n-k+1}a_{n-1}^{s_1}\cdots
a_{n-k}^{s_{k}}+c_na_{n-1}^{m-1}\\
             &=\sum\limits_{k=1}^{m}\sum\limits_{s_1+s_2+\cdots
+s_{k}=m-k}c_n\cdots c_{n-k+1}a_{n-1}^{s_1}\cdots
a_{n-k}^{s_{k}}\\
             &=\sum\limits_{k=1}^{m}\sum\limits_{\substack{s_0+s_1+\cdots
+s_{k}=m-k\\ s_0=0}}c_n\cdots c_{n-k+1}a_n^{s_0}a_{n-1}^{s_1}\cdots
a_{n-k}^{s_{k}}\\
\end{split}
\end{displaymath}
From the above two equation, there holds\\
\begin{displaymath}
\begin{split}
f(m,n)&=a_nf(m-1,n)+c_nf(m-1,n-1)\\
      &=\sum\limits_{k=1}^{m}\sum\limits_{\substack{s_0+s_1+\cdots
      +s_k=m-k\\s_0,\cdots s_k\geq 0}}c_n\cdots c_{n-k+1}a_{n}^{s_0}\cdots
a_{n-k}^{s_k}+a_{n}^{m}
\end{split}
\end{displaymath}
The previous $n$ can be replaced by any $l$, where $m\leq l\leq n$
$\Box$\\

Using lemma1 we can easily obtain two following Theorems.\\\\
\textbf{Theorem3.} Choose $A_{m,n}$ u.a.r. from $\mathcal{A}(m,n)$, $m\leq n$, or equivalently let $A_{m,n}$ from $\mathcal {A}(m,n,1/2)$. Then\\
\begin{displaymath}
E_{\mathcal{A}}(AM(A_{m,n}))=\sum\limits_{k=0}^{m}C_m^k\frac{P_n^{k}}{2^{k}}
\end{displaymath}
where $C_m^k=\frac{m!}{k!(m-k)!}$ and $P_n^k=\frac{n!}{(n-k)!}$\\
\textbf{Proof:} Induction on m. The case p=0,
$E_{\mathcal{A}}(AM(A))=1$ is trivial. Suppose $\forall$ $p\leq
m-1$, $p\leq l\leq n$
$E_{\mathcal{A}}(AM(A_{p,l}))=\sum\limits_{k=0}^{p}C_p^k\frac{P_l^{p-k}}{2^{p-k}}
=\sum\limits_{k=0}^{p}C_p^k\frac{P_l^{k}}{2^{k}}$\\
when p=m, $\forall$ $m\leq l\leq n$, we have\\
\begin{displaymath}
\begin{split}
E_{\mathcal{A}}(AM(A_{m,l}))&=E_{\mathcal{A}}(AM(A_{m,l}^{1,0})+\sum\limits_{j=1}^{n}a_{1,j}AM(A_{m,l}^{1,j}))\\
                            &=E_{\mathcal{A}}(AM(A_{m-1,l}))+\sum\limits_{j=1}^{n}E_{\mathcal{A}}(a_{1,j})E_{\mathcal{A}}(AM(A_{m-1,l-1})\\
                            &=E_{\mathcal{A}}(AM(A_{m-1,l}))+\frac{n}{2}E_{\mathcal{A}}(AM(A_{m-1,l-1})\\
\end{split}
\end{displaymath}
Using lemma1, here $a_l\equiv 1$, and$c_l=\frac{l}{2} $ then
\begin{displaymath}
\begin{split}
E_{\mathcal{A}}(AM(A_{m,l}))&=\sum\limits_{k=1}^{m}\sum\limits_{\substack{s_0+s_1+\cdots
+s_k=m-k \\s_0,\cdots s_k\geq 0}}c_l\cdots c_{l-k+1}+1\\
                            &=\sum\limits_{k=1}^{m}\frac{P_l^k}{2^k}\sum\limits_{\substack{s_0+s_1+\cdots
+s_k=m-k\\ s_0,\cdots s_k\geq 0}}1+1\\
                            &=\sum\limits_{k=1}^{m}\frac{P_l^k}{2^k}C_m^k+1\\
                            &=\sum\limits_{k=0}^{m}\frac{P_l^k}{2^k}C_m^k
\end{split}
\end{displaymath}
$\Box$ \\
\textbf{Theorem4} Choose $A_{m,n}$ u.a.r. from $\mathcal{A}(m,n)$,
$m\leq n$, and let $X_{A_{m,n}}$ be the output by AMM. Then
\begin{displaymath}
E_{\mathcal{A}}(E_{\sigma}(X_{A_{m,n}}))=\sum\limits_{k=0}^{m}C_m^k\frac{P_n^{k}}{2^{k}}
\end{displaymath}
and
\begin{displaymath}
E_{\mathcal{A}}(E_{\sigma}(X_{A_{m,n}}^2))=\sum\limits_{k=0}^{m}\frac{P_n^kP_{n+3}^{k}}{2^{m+k}}\sum\limits_{\substack{s_0+s_1+\cdots
+s_k=m-k\\ s_0,\cdots s_k\geq 0}}(n+2)^{s_0}(n+2-1)^{s_1}\cdots
(n+2-k)^{s_k}
\end{displaymath}
\textbf{Proof:} The first equation is is trivial since
$E_{\sigma}(X_{A_{m,n}}^2)=AM(A_{m,l})$. For the second one, we use
induction on m. The case p=0 is obvious. Suppose $\forall A_{p,l}$
where $0\leq p\leq m-1$, $p\leq l\leq n$ the second equation holds.
When $p=m$, noting the fact $M=|W|-1$ is a binomial variable with
parameter $l$ and $1/2$(recall $W/\{0\}$ is the set of column
indices with a 1 in the first row), then
\begin{displaymath}
\begin{split}
E_{\mathcal{A}}(E_{\sigma}(X_{A_{m,l}}^2))&=\sum\limits_{q=0}^{l}E_{\mathcal{A}}(E_{\sigma}(X_{A_{m,l}}^2)|M=q)Pr_{\mathcal{A}}(M=q)\\
                                          &=\sum\limits_{q=0}^{l}E_{\mathcal{A}}((q+1)\sum\limits_{j\in W}E_{\sigma}(X_{A_{m,l}^{1j}}^2)|M=q)Pr_{\mathcal{A}}(M=q)\\
                                          &=\sum\limits_{q=0}^{l}E_{\mathcal{A}}((q+1)E_{\sigma}(X_{A_{m-1,l}}^2)+q(q+1)E_{\sigma}(X_{A_{m-1,l-1}}^2))Pr_{\mathcal{A}}(M=q)\\
                                          &=(E_{\mathcal{A}}(M)+1)E_{\mathcal{A}}(E_{\sigma}(X_{A_{m-1,l}}^2))+(E_{\mathcal{A}}(M^2)+E_{\mathcal{A}}(M))E_{\mathcal{A}}(E_{\sigma}(X_{A_{m-1,l-1}}^2))\\
                                          &=(\frac{l+2}{2})E_{\mathcal{A}}(E_{\sigma}(X_{A_{m-1,l}}^2))+(\frac{l^2+3l}{4})E_{\mathcal{A}}(E_{\sigma}(X_{A_{m-1,l-1}}^2))\\
\end{split}
\end{displaymath}
Using lemma1, here $a_l=\frac{l+2}{2}$, and
$c_l=\frac{l^2+3l}{4}$.Then
\begin{displaymath}
\begin{split}
E_{\mathcal{A}}(E_{\sigma}(X_{A_{m,l}}^2))&=\sum\limits_{k=1}^{m}\frac{P_l^kP_{l+3}^{k}}{4^{k}}\sum\limits_{\substack{s_0+s_1+\cdots
+s_k=m-k \\s_0,\cdots s_k\geq 0}}(\frac{l+2}{2})^{s_0}(\frac{l+2-1}{2})^{s_1}\cdots (\frac{l+2-k}{2})^{s_k}+(\frac{l+2}{2})^m\\
                                          &=\sum\limits_{k=0}^{m}\frac{P_l^kP_{l+3}^{k}}{2^{k+m}}\sum\limits_{\substack{s_0+s_1+\cdots
+s_k=m-k \\s_0,\cdots s_k\geq 0}}(l+2)^{s_0}(l+2-1)^{s_1}\cdots (l+2-k)^{s_k}\\
\end{split}
\end{displaymath}
$\Box$\\
\textbf{Theorem5} Choose $A_{n,n}$ u.a.r. from $\mathcal{A}(n,n)$,
and let $X_{A_{n,n}}$ be the output by AMM. Then \textbf{whp} $h(n)
\leq E_{\mathcal{A}}(E_{\sigma}(X_{A_{n,n}}))\leq nh(n)$, where
$h(n)=\frac{(n!)^2}{2^n}\frac{2^{k^{\ast}}}{(n-k^{\ast})!(k^{\ast})^2}$,
$k^{\ast}=\lfloor -1+\sqrt{2n+3}\rfloor$. where $\lfloor\ast\rfloor$
denotes the largest integer no more than $\ast$.\\
\textbf{Proof:}
\begin{displaymath}
\begin{split}
E_{\mathcal{A}}(E_{\sigma}(X_{A_{n,n}}))&=\sum\limits_{k=0}^{n}C_n^k\frac{P_n^{k}}{2^{k}}\\
&=\sum\limits_{k=0}^{n}C_n^k\frac{P_n^{n-k}}{2^{n-k}}\\
&=\frac{(n!)^2}{2^n}\sum\limits_{k=0}^{n}\frac{2^k}{(n-k)!(k!)^2}
\end{split}
\end{displaymath}
and let $b_k=\frac{2^k}{(n-k)!(k!)^2}$, then
$\frac{b_k}{b_{k-1}}=\frac{2(n-k+1)}{k^2}$, set
$\frac{b_k}{b_{k-1}}\geq 1$ we have $k\leq -1+\sqrt{2n+3}$, thus,
$b_{k^{\ast}}=\max\limits_{k=0,\cdots,n}b_k$. Thus, obviously\\
$\frac{(n!)^2}{2^n}b_{k^{\ast}}\leq
E_{\mathcal{A}}(E_{\sigma}(X_{A_{n,n}}))\leq
n\frac{(n!)^2}{2^n}b_{k^{\ast}}$\\
$\Box$ \\
\textbf{Theorem6} Choose $A_{n,n}$ u.a.r. from $\mathcal{A}(n,n)$,
and let $X_{A_{n,n}}$ be the output by AMM. Then \textbf{whp}
\begin{displaymath}
\frac{E_{\mathcal{A}}(E_{\sigma}(X_{A_{n,n}}^2))}{E_{\mathcal{A}}^2(E_{\sigma}(X_{A_{n,n}}))}
\geq n^{(\sqrt{n}/2)}
\end{displaymath}
\textbf{Proof:} Numerical experiment shows the above result. however
the theoretical analysis seems so hard than until now I haven't
thought out the way to show the comparably tight for
$E_{\mathcal{A}}(E_{\sigma}(X_{A_{n,n}}^2))$ since the order of
$\sum\limits_{s_0+s_1+\cdots +s_k=n-k}(n+2)^{s_0}(n+2-1)^{s_1}\cdots
(n+2-k)^{s_k}$ is too difficult to gain a good lower bound. The
following bound is easy to check and the best one among methods I
thought out,
\begin{displaymath}
E_{\mathcal{A}}(E_{\sigma}(X_{A_{n,n}}^2))\geq
\sum\limits_{k=0}^{n}\frac{(n!)^2(n+3)!}{2^{2n}}\frac{2^k(k+2)^k}{(k!)^2(k+3)!(n-k)!}
\end{displaymath}
However it still can't reach the goal. Therefore, the proof of this
theorem will be left for the future.

Even if Theorem6 has been proved, unfortunately, the critical ratio
for almost all the matrices can not obtained from this theorem since
two random variables are not independent. In order to accomplish the
ultimate result, we need to calculate the
$E_{\mathcal{A}}(E_{\sigma}^2(X_{A_{n,n}}^2))$. Using the induction
similar to theorem4, we can obtain the recursion of
$E_{\mathcal{A}}(E_{\sigma}^2(X_{A_{m,n}}^2))$(recall M is a
binomial variable with parameter n and $\frac{1}{2}$).
\begin{multline*}
E_{\mathcal{A}}(E_{\sigma}^2(X_{A_{m,n}}^2))=
2(E_{\mathcal{A}}(M^3)+2E_{\mathcal{A}}(M^2)+E_{\mathcal{A}}(M))E_{\mathcal{A}}(E_{\sigma}(X_{A_{m-1,n}}^2)E_{\sigma}(X_{A_{m-1,n-1}}^2))\\
+(E_{\mathcal{A}}(M^2)+2E_{\mathcal{A}}(M)+1)E_{\mathcal{A}}(E_{\sigma}^2(X_{A_{m-1,n}}^2))
+(E_{\mathcal{A}}(M^4)+2E_{\mathcal{A}}(M^3)+E_{\mathcal{A}}(M^2))E_{\mathcal{A}}(E_{\sigma}^2(X_{A_{m-1,n-1}}^2))
\end{multline*}
Comparing $E_{\mathcal{A}}(E_{\sigma}^2(X_{A_{m,n}}^2))$ with
$E_{\mathcal{A}}^2(E_{\sigma}(X_{A_{n,n}}^2))$ and computing their
ratio have to be done. Our main aim of doing this is to find the
matrices satisfying $E_{\sigma}(X_{A_{m,n}}^2)\leq
E_{\mathcal{A}}(E_{\sigma}^2(X_{A_{m,n}}^2))g(n)$, where $g(n)$ is a
polynomial of $n$. However, the ratio of
$\frac{E_{\mathcal{A}}(E_{\sigma}^2(X_{A_{m,n}}^2))}{E_{\mathcal{A}}^2(E_{\sigma}(X_{A_{n,n}}^2))}$
is so large that it can't accomplish our goal. Thus we deduce our
requirement \textbf{whp} to with a certain probability $p>0$, and in
our results $p=\frac{1}{2}-\varepsilon$ where $\varepsilon$ is no
more than 0.02. To prove the theorem, we need the following
lemma, which will be proved in section IV.\\
\textbf{Lemma2} Let $\mathcal {B}(m,n)$ denote the set of all
$n\times n$ 0-1 matrices with exact m 1's, $m \gg n$. Choose $B$
u.a.r. from $\mathcal {B}(m,n)$. Then
\begin{displaymath}
E(AM(B))=\sum\limits_{k=0}^{n}(C_n^k)^2k!\frac{C_{n^2-k}^{m-k}}{C_{n^2}^{m}}
\end{displaymath}
and
\begin{displaymath}
\frac{E(AM^2(B))}{E^2(AM(B))}=1+o(1), n\rightarrow \infty
\end{displaymath}
\textbf{Theorem7} Choose $A_{n,n}$ u.a.r. from $\mathcal{A}(n,n)$,
and let $X_{A_{n,n}}$ be the output by AMM. Then
\begin{displaymath}
Pr(\frac{E_{\sigma}(X_{A_{n,n}}^2)}{E_{\sigma}^2(X_{A_{n,n}})}\geq
n^{\sqrt{n}/2-1})\geq\frac{\sum\limits_{i=(1/2+\varepsilon)n^2}^{n^2}C_{n^2}^k}{2^{n^2}}
\end{displaymath}
where c is a constant no more 10, and$\varepsilon\leq 0.02$.\\
\textbf{Proof:} From lemma2 we know if we set
$m=(1/2+\varepsilon)n^2$ and
$q=\frac{C_{n^2-k}^{m-k}}{C_{n^2}^{m}}$. When n goes to infinity,
noting $k\leq n\ll m, n^2$, there holds
\begin{displaymath}
q=\frac{C_{n^2-k}^{m-k}}{C_{n^2}^{m}}=\frac{m(m-1)\cdots
(m-k)}{n^2(n^2-1)\cdots (n^2-k)}
\end{displaymath}
and
\begin{displaymath}
\begin{split}
ln(q)&=\sum\limits_{i=0}^{k-1}[ln(m-i)-ln(n^2-i)]\\
&=kln(\frac{m}{n^2})+\sum\limits_{i=0}^{k-1}[ln(1-\frac{i}{m})-ln(1-\frac{i}{n^2})]\\
&=kln(\frac{m}{n^2})-\sum\limits_{i=0}^{k-1}[\frac{i}{m}-\frac{i}{n^2}+O(\frac{i^2}{m^2})]\\
&=kln(\frac{m}{n^2})-\frac{k(k-1)}{2}(\frac{1}{m}-\frac{1}{n^2})+O(\frac{k^3}{m^2})
\end{split}
\end{displaymath}
Thus, noting that $km^{-1}\leq 2nm^{-1}=O(n^3m^{-2})$
\begin{displaymath}
\begin{split}
q&=(\frac{m}{n^2})^kexp[-\frac{k^2}{2}(\frac{1}{m}-\frac{1}{n^2})+O(\frac{n^3}{m^2})]\\
&=(\frac{(1/2+\varepsilon)n^2}{n^2})^kexp[-\frac{k^2}{2}(\frac{1}{(1/2+\varepsilon)n^2}-\frac{1}{n^2})+O(\frac{n^3}{((1/2+\varepsilon)n^2)^2})]\\
&\leq e^{-1}(1/2+\varepsilon)^k
\end{split}
\end{displaymath}
Let B selected u.a.r. from $\mathcal {B}(m,n)$ Since
$\frac{E(AM^2(B))}{E^2(AM(B))}=1+o(1)$, as $n\rightarrow \infty$\\
then $Pr(AM(B)<\frac{5}{6}E(AM(B)))\rightarrow 0$, as $n\rightarrow
\infty$. So, if $m\geq (1/2+\varepsilon)n^2$ and $\varepsilon\leq
0.02$, we have \textbf{whp}
\begin{displaymath}
\begin{split}
E_{\sigma}(X_{B}^2)&\geq E_{\sigma}^2(X_{B})\\
&= AM^2(B)\\
&\geq (\frac{5}{6}E(AM(B)))^2 \\
&=(\frac{5}{6}\sum\limits_{k=0}^{n}(C_n^k)^2k!\frac{C_{n^2-k}^{m-k}}{C_{n^2}^{m}})^2\\
&\geq(\sum\limits_{k=0}^{n}(C_n^k)^2k!\frac{5e^{-1}}{6}(1/2+\varepsilon)^k)^2\\
&\geq\sum\limits_{k=0}^{n}\frac{P_n^kP_{n+3}^{k}}{2^{n+k}}\sum\limits_{\substack{s_0+s_1+\cdots
+s_k=n-k \\s_0,\cdots s_k\geq 0}}(n+2)^{s_0}(n+2-1)^{s_1}\cdots
(n+2-k)^{s_k}\\
&=E_{\mathcal{A}}(E_{\sigma}(X_{A_{n,n}})).
\end{split}
\end{displaymath}
Noting $Pr(A\in\bigcup\limits_{m\geq (1/2+\varepsilon)n^2}\mathcal
{B}(m,n))=\frac{\sum\limits_{i=(1/2+\varepsilon)n^2}^{n^2}C_{n^2}^k}{2^{n^2}}$,\\
thus $Pr(E_{\sigma}(X_{A_{n,n}}^2)\geq
E_{\mathcal{A}}(E_{\sigma}(X_{A_{n,n}})))\geq \frac{\sum\limits_{i=(1/2+\varepsilon)n^2}^{n^2}C_{n^2}^k}{2^{n^2}}$.\\
Using Markov's inequality,
\begin{displaymath}
Pr(E_{\sigma}(X_{A_{n,n}}^2)\geq n
E_{\mathcal{A}}(E_{\sigma}(X_{A_{n,n}})))\leq \frac{1}{n}
\rightarrow 0
\end{displaymath}
then \textbf{whp} $E_{\sigma}(X_{A_{n,n}}^2)\leq n
E_{\mathcal{A}}(E_{\sigma}(X_{A_{n,n}}))$. Finally, we have
\begin{displaymath}
Pr(\frac{E_{\sigma}(X_{A_{n,n}}^2)}{E_{\sigma}^2(X_{A_{n,n}})}\geq
\frac{1}{n}\frac{E_{\mathcal{A}}(E_{\sigma}(X_{A_{n,n}}^2))}{E_{\mathcal{A}}(E_{\sigma}(X_{A_{n,n}}))})\geq
\frac{\sum\limits_{i=(1/2+\varepsilon)n^2}^{n^2}C_{n^2}^k}{2^{n^2}}
\end{displaymath}
Apply theorem6 to the above formula, we have
\begin{displaymath}
Pr(\frac{E_{\sigma}(X_{A_{n,n}}^2)}{E_{\sigma}^2(X_{A_{n,n}})}\geq
n^{\sqrt{n}/2-1})\geq
\frac{\sum\limits_{i=(1/2+\varepsilon)n^2}^{n^2}C_{n^2}^k}{2^{n^2}}
\end{displaymath}

\section {The number of all the matchings on random graph.}

In this section, we consider the expectation and variance of the
number of all the matchings on G selected u.a.r. from $\mathcal
{G}(m,n)$. We have the following theorem. \\\\
\textbf{Theorem8} Choose G u.a.r. from $\mathcal {G}(m,n)$, where
$\mathcal {G}(m,n)$ denotes the set of bipartite graph with
$\#V_1=\#V_2=n$ as its vertices and exact $m$ edges, $m \gg n$, and
let AM(G) denotes the number of all the matchings in G. Then we have
\begin{displaymath}
E(AM(G))=\sum\limits_{k=0}^{n}(C_n^k)^2k!E(X_{M(k)})
\end{displaymath}
and
\begin{multline*}
E(AM^2(G))=\sum\limits_{k=0}^{n}\sum\limits_{i=0}^{k}(C_n^k)^2k!\sum\limits_{p=0}^{min(i,n-k)}
C_{n-k}^{p}C_{k}^{i-p}P_{n-i+p}^{p}\sum\limits_{j=0}^{i-p}C_{i-p}^{j}[F_{n-j}(i-p-j)]E(X_{M(k+i-j)})\\
+\sum\limits_{k=1}^{n}\sum\limits_{i=0}^{k-1}(C_n^k)^2k!\sum\limits_{p=0}^{min(i,n-k)}
C_{n-k}^{p}C_{k}^{i-p}P_{n-i+p}^{p}\sum\limits_{j=0}^{i-p}C_{i-p}^{j}[F_{n-j}(i-p-j)]E(X_{M(k+i-j)})
\end{multline*}
where $E(X_{M(k)})=C_{n^2-k}^{m-k}/C_{n^2}^{m}$ and
$F_{n}(p)=\sum\limits_{r=0}^{p}(-1)^rC_p^rP_{n-r}^{p-r}$\\
\textbf{Proof:} we'll use the methodology in \cite{Jerr95}; Let
$M(k)$ be a k-matching on $V_{1}+V_{2}$, For $G\in\mathcal
{G}(m,n)$, define the random variable $X_M(G)$ to be 1 if $M(k)$ is
contained in G, and otherwise 0.  The expectation and second moment
of $AM(G)$ is as follows. \begin{displaymath}
E(AM(G))=E(\sum\limits_{k=0}^{n}\sum\limits_{M(k)}X_{M(k)})=\sum\limits_{k=0}^{n}\sum\limits_{M(k)}E(X_{M(k)})
\end{displaymath}
and
\begin{displaymath}
E(AM^2(G))=E((\sum\limits_{k=0}^{n}\sum\limits_{M(k)}X_{M(k)})^2)=\sum\limits_{k=0}^{n}\sum\limits_{i=0}^{n}\sum\limits_{M(k),M^{'}(i)}E(X_{M(k)}X_{M(i)}^{'})
\end{displaymath}
where $\forall 0\leq k\leq n$, $M(k)$ and $M^{'}(k)$ range over all
$(C_{n}^{k})^2k!$ k-matching's on $V_{1}+V_{2}$. Note that
\begin{displaymath}
E(X_{M(k)})=\frac{C_{n^2-k}^{m-k}}{C_{n^2}^{m}}
\end{displaymath}
The first equation follows quickly. For the second, in order to
compute $E(X_{M(k)}X_{M(i)}^{'})$, we have to calculate the number
of pairs of $M(k)$ and $M^{'}(i)$ as a function of the overlap
$j=|M(k)\bigcap M^{'}(i)|$. For any fixed $k$, suppose $i\leq k$, we
need to compute the number of the pairs of $M(k)$ and $M^{'}(i)$,
where $i=0,\cdots,k$, and $M^{'}(i)$ ranges over all
$(C_{n}^{i})^2i!$ $i$-matching's on $V_{1}+V_{2}$. The problem can
be equivalently stated as follows: There're $n$ different letters
and $n$ different envelopes. Among these letters, there're exact
$k(0\leq k\leq n)$ labeled letters, each of which has only one
'mother envelope' among envelopes. Different labeled letters have
different mother envelopes. We call a $j$-fit if there're exact $j$
labeled letters put into its own mother envelope. Now choose
$i$($0\leq i\leq k$)letters from these $n$ letters, then put them
into $i$ envelopes, and each letter can only be put into one
envelope. $\forall$ possible $j$, how many circumstances of $j$-fit
are there? We can solve this problem like this: Suppose there're $p$
letters unlabeled and $i-p$ labeled letters among the selected
letters, obviously, $0\leq p\leq min(n-k,i)$, the number of ways of
choosing letters is $C_{n-k}^{p}C_{k}^{i-p}$. If the labeled letters
has been laid, then the number of the ways of putting $p$ unlabeled
letters is $P_{n-(i-p)}^{p}$. For any $j$($0\leq j\leq i-p$),
there're $C_{i-p}^{j}$ ways putting exact $j$ labeled letters in its
own mother envelope. The last one we need to deal with is how many
ways to put $i-p-j$ labeled letters into $n-j$ envelopes which
contain all these $i-p-j$ letters' mother envelopes, satisfying
$0$-fit. By the principle of inclusion-exclusion see\cite{Hal67}, we
can easily obtain the number of the ways is $F_{n-j}(i-p-j)$, where
$F_{n}(p)=\sum\limits_{r=0}^{p}(-1)^rC_p^rP_{n-r}^{p-r}$. Noting
that $p$ ranges over $0$ to $min(i,n-k)$, and $j$ ranges over $0$ to
$i-p$, for each $k$ and $i\leq k$. Then
\begin{displaymath}
\begin{split}
\sum\limits_{M^{'}(i)}E(X_{M(k)}X_{M(i)}^{'})&=\sum\limits_{p=0}^{min(i,n-k)}
C_{n-k}^{p}C_{k}^{i-p}P_{n-i+p}^{p}\sum\limits_{j=0}^{i-p}C_{i-p}^{j}[F_{n-j}(i-p-j)]E(X_{M(k+i-j)})
\end{split}
\end{displaymath}
where $E(X_{M(k)})=C_{n^2-k}^{m-k}/C_{n^2}^{m}$ and
$F_{n}(p)=\sum\limits_{r=0}^{p}(-1)^rC_p^rP_{n-r}^{p-r}$.\\
Consider,
\begin{displaymath}
\begin{split}
\sum\limits_{k=0}^{n}\sum\limits_{i=0}^{n}\sum\limits_{M(k),M^{'}(i)}E(X_{M(k)}X_{M(i)}^{'})
&=(\sum\limits_{k=0}^{n}\sum\limits_{i=0}^{k}+
\sum\limits_{k=0}^{n-1}\sum\limits_{i=k+1}^{n})\sum\limits_{M(k),M^{'}(i)}E(X_{M(k)}X_{M(i)}^{'})\\
&=(\sum\limits_{k=0}^{n}\sum\limits_{i=0}^{k}+
\sum\limits_{k=0}^{n-1}\sum\limits_{i=k+1}^{n})\sum\limits_{M(k),M^{'}(i)}E(X_{M(k)}X_{M(i)}^{'})\\
&=(\sum\limits_{k=0}^{n}\sum\limits_{i=0}^{k}+
\sum\limits_{i=1}^{n}\sum\limits_{k=0}^{i-1})\sum\limits_{M(k),M^{'}(i)}E(X_{M(k)}X_{M(i)}^{'})\\
&=(\sum\limits_{k=0}^{n}\sum\limits_{i=0}^{k}+
\sum\limits_{k=1}^{n}\sum\limits_{i=0}^{k-1})\sum\limits_{M(k),M^{'}(i)}E(X_{M(k)}X_{M(i)}^{'})\\
&=(\sum\limits_{k=0}^{n}\sum\limits_{i=0}^{k}+
\sum\limits_{k=1}^{n}\sum\limits_{i=0}^{k-1})\sum\limits_{M(k)}\sum\limits_{M^{'}(i)}E(X_{M(k)}X_{M(i)}^{'})\\
&=(\sum\limits_{k=0}^{n}(C_{n}^{k})^2k!\sum\limits_{i=0}^{k}+
\sum\limits_{k=1}^{n}(C_{n}^{k})^2k!\sum\limits_{i=0}^{k-1})\sum\limits_{M^{'}(i)}E(X_{M(k)}X_{M(i)}^{'})
\end{split}
\end{displaymath}
Replace $\sum\limits_{M^{'}(i)}E(X_{M(k)}X_{M(i)}^{'})$ by\\
$\sum\limits_{p=0}^{min(i,n-k)}
C_{n-k}^{p}C_{k}^{i-p}P_{n-i+p}^{p}\sum\limits_{j=0}^{i-p}C_{i-p}^{j}[F_{n-j}(i-p-j)]E(X_{M(k+i-j)})$,
then the second equation is achieved.\\
$\Box$\\

\textbf{Remark:} To complete the proof of theorem7, we also need to
know whether the ratio $\frac{E(AM^2(G))}{E^2(AM(G))}$ goes to 1 as
n goes to infinity, adding the condition such as $m^{2}n^{-3}$
$\rightarrow$ $\infty$ as $n$ $\rightarrow$ $\infty$. We guess such
a result is right, however the calculus seems very difficult. And
this result also contributes to the upper bound of critical ratio
for almost all the matrices.
\begin{acknowledgements}
\end{acknowledgements}
{\baselineskip=12pt \small{
}
\end{document}